\documentclass[%
reprint,
superscriptaddress,
amsmath,
amssymb,
aps,
rmp,
floatfix,
]{revtex4-2}

\usepackage{graphicx}
\usepackage{dcolumn}
\usepackage{bm}
\usepackage{hyperref}
\usepackage[mathlines]{lineno}

\usepackage{stmaryrd} 
\usepackage{xcolor, soul}
\usepackage{placeins} 
\usepackage{subcaption} 
\usepackage{enumerate} 
\usepackage{yhmath} 

\begin{document}
\preprint{APS/123-QED}

\title{Stylized facts in money markets: an empirical analysis of the eurozone data}

\author{Victor Le Coz}
\email{victor.lecoz@gmail.com}
\affiliation{Quant AI Lab, 29 Rue de Choiseul 75002 Paris, France}
\affiliation{Chair of Econophysics and Complex Systems, \'Ecole polytechnique, 91128 Palaiseau Cedex, France}
\affiliation{LadHyX UMR CNRS 7646, \'Ecole polytechnique, 91128 Palaiseau Cedex, France}
\affiliation{Laboratoire de Math\'ematiques et Informatique pour la Complexit\'e et les Syst\`emes, CentraleSupélec, Universit\'e Paris-Saclay, 91192 Gif-sur-Yvette Cedex, France}

\author{Nolwenn Allaire}
\email{nolwenn.allaire@ecb.europa.eu}
\affiliation{European Central Bank, Sonnemastrasse 20, 60314 Frankfurt am Main, Germany}

\author{Michael Benzaquen}
\email{michael.benzaquen@polytechnique.edu}
\affiliation{Chair of Econophysics and Complex Systems, \'Ecole polytechnique, 91128 Palaiseau Cedex, France}
\affiliation{LadHyX UMR CNRS 7646, \'Ecole polytechnique, 91128 Palaiseau Cedex, France}
\affiliation{Capital Fund Management, 23 Rue de l’Universit\'e, 75007 Paris, France}

\author{Damien Challet}
\email{damien.challet@centralesupelec.fr}
\affiliation{Laboratoire de Math\'ematiques et Informatique pour la Complexit\'e et les Syst\`emes, CentraleSupélec, Universit\'e Paris-Saclay, 91192 Gif-sur-Yvette Cedex, France}

\date{\today}

\begin{abstract}
Using the secured transactions recorded within the Money Markets Statistical Reporting database of the European Central Bank, we test several stylized facts regarding interbank market of the $47$-largest banks in the eurozone. We observe that the surge in the volume of traded evergreen repurchase agreements followed the introduction of the LCR regulation and we measure a rate of collateral re-use consistent with the literature. Regarding the topology of the interbank network, we confirm the high level of network stability but observe a higher density and a higher in-- and out--degree symmetry than what is reported for unsecured markets.
\end{abstract}

\maketitle


\section{Introduction}
Money markets are venues where banks carry out their refinancing activities. The 2008 Great Financial Crisis (GFC) led to heightened counterparty risk and, consequently, significant changes in money markets across Western economies. In response, the European Central Bank (ECB) introduced the full allotment procedure in October 2008, allowing banks to access unlimited central bank financing. Meanwhile, the Basel regulations mandated the Liquidity Coverage Ratio (LCR) to improve banks' short-term liquidity resilience, requiring them to hold a sufficient amount of high-quality liquid assets to meet their liquidity needs during stress scenarios. These rules resulted in the creation of excess reserves within the financial system \citep{Renne-2012,PiquardSalakhova-2019,LucaBaldoEtAl-2017}. Additionally, the banking system's refinancing increasingly relied on collateralized lending, particularly through repurchase agreements (\textit{repos}), with a notable rise in the practice of collateral re-use \citep{KellerEtAl-2014,EuropeanSystemicRiskBoard.-2017,CheungEtAl-2014,FuhrerEtAl-2016,SCAGGS-2018,Accornero-2020}.

The network topology of money markets, where transactions among banks are identified as links between nodes, has evolved in consequence but was not yet documented. This study aims at quantifying the changes of the secured funding operations of the 47 largest banks in the eurozone.

\section{Excess liquidity and declining unsecured interbank markets}

\begin{figure}
\centering
\includegraphics[width=0.5\textwidth]{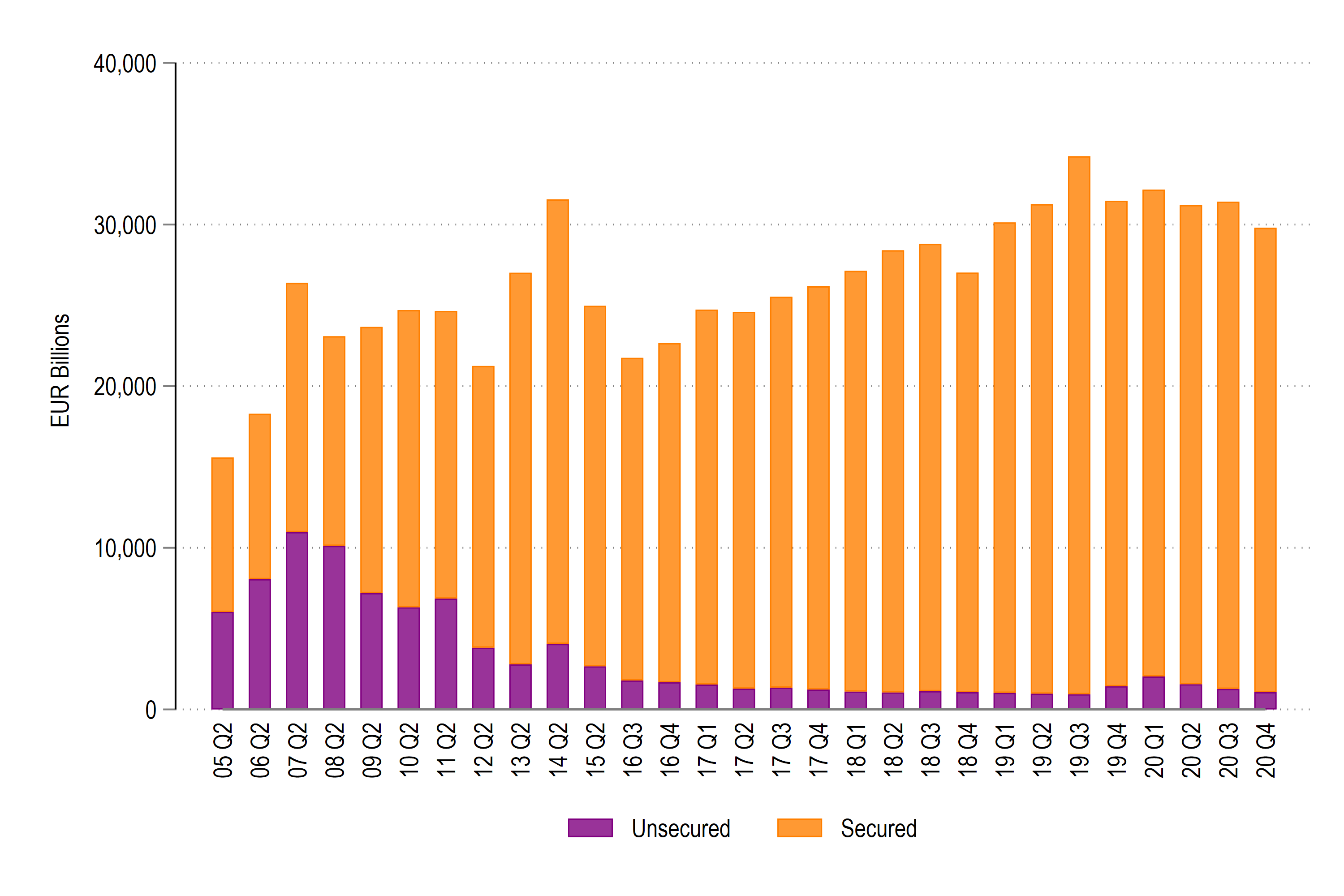}
\caption*{\footnotesize{Cumulative quarterly turnover in the euro area unsecured and secured money market segments. Source: Euro Area Money Market Survey until Q2 2015, Money Market Statistical Repoting (MMSR) data thereafter. Only transactions with deposit-taking institutions and CCPs are considered. Both borrowing and lending transactions are included; all collateral types and maturities are considered.}}
\caption{Turnover in unsecured and secured euro area interbank money markets}  
\label{fig:MMSR_Turnover_Motivation_1}
\end{figure}

Following the GFC, the ECB introduced the full allotment procedure, enabling banks to meet any liquidity requirements without limitation \citep{Renne-2017}. Consequently, volumes in the overnight unsecured interbank market declined markedly (see Fig.~\ref{fig:MMSR_Turnover_Motivation_1}). As noted in a recent ECB survey \citep{LucaBaldoEtAl-2017}, the increase in excess liquidity between 2012 and 2018 can be attributed to three main factors: (i) heightened demand for central bank liquidity from banks, (ii) the implementation of the full allotment procedure, and (iii) the provision of longer-term refinancing operations. Since 2015, the ECB's asset purchase program (APP) further augmented excess liquidity in the banking system, with most banks reporting that client inflows were a predominant factor \citep{LucaBaldoEtAl-2017}. The subsequent decrease in unsecured lending was exacerbated by the introduction of the LCR in January 2018, which impeded liquidity redistribution \citep{LucaBaldoEtAl-2017}. Indeed, \citet{LeCozEtAl-2024c} demonstrated, through accounting analysis, that the interaction between the APP and the LCR requirement contributes to the persistence of excess liquidity in the financial system.

\section{Evergreen repos to answer LCR regulation} \label{Creation of evergreen repos to answer to the LCR regulation}

The GFC underscored the presence of counterparty risk among banks, prompting a shift from unsecured to secured lending \citep{FilippoEtAl-2018}. In these markets, collateralized borrowings are executed through repurchase agreements (repos), which involve the exchange of collateral for cash over a specified period. This shift towards secured markets was further reinforced by the introduction of the LCR, as repos can effectively navigate this regulatory constraint. An evergreen repo, a contract that is continually renewed by mutual agreement, exemplifies this adaptation. \citet{LeCozEtAl-2024c} demonstrate that an evergreen repo with a one-month notice period has no impact on the LCR of the parties involved, as the collateral provided offsets any LCR loss due to the cash exchange.

\begin{figure}
\centering
\includegraphics[width=0.5\textwidth]{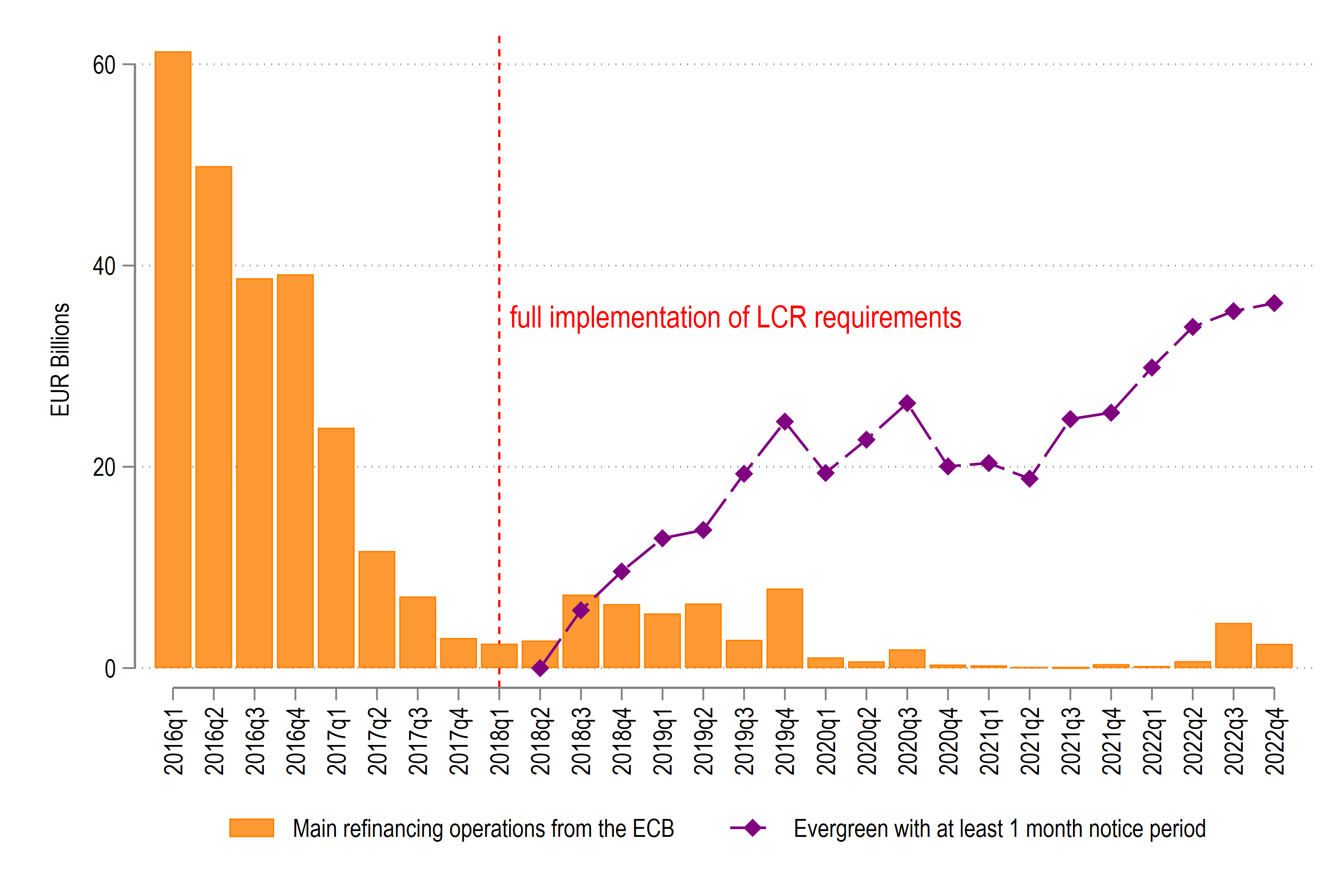} 
\caption*{\footnotesize{Aggregate quarterly volumes of main refinancing operations (MRO) implemented by the Eurosystem and aggregate volumes of traded evergreen repos with a notice period greater than 1 month. Source: Internal Liquidity Management for the MRO volumes and Money Market Statistical Reporting (MMSR) data for evergreen repos. Evergreen repos are identified by filtering on repo transactions with a notice period of at least 30 days, with repeating transactions for at least 1 day. Both borrowing and lending transactions are included; all collateral types are considered.}}
\caption{Evergreen repos}  
\label{fig:MMSR_Evergreen_Motivation_2}
\end{figure}

Figure~\ref{fig:MMSR_Evergreen_Motivation_2} shows that the introduction of the LCR regulation coincides with an increase in the volumes of traded evergreen repos with a notice period longer than one month (see also \citet{Allen-2016}).  We observe that the volume of evergreen repos traded among the $47$ largest banks in the eurozone increased from negligible amounts in $2017$ to ten billions per day in $2019$. All the empirical results presented here were established for the $47$ largest banks in the eurozone that are required to report their transactions to the Money Market Statistical Reporting database (MMSR) as detailed in appendix~\ref{Data sources}.

\section{Collateral re-use and bond scarcity}

The one-month notice period of evergreen repos prevents the immediate unwinding of positions when the cash lender faces a liquidity requirement. This constraint facilitates the re-use of collateral in the market. Specifically, the cash lender $j$ can re-use the collateral $S^c_{j,t}$ received from a reverse repo to secure cash in another repo transaction. The rate at which collateral is re-used has been defined in various ways within the literature \citep{Accornero-2020}. Here, we define the re-use rate of collateral as:
\begin{align}
    \text{re-use}(t) = \frac{\sum_{i=1}^NS^{r}_{i}(t)}{\sum_{i=1}^NS^{c}_{i}(t)}.
\end{align}

\begin{figure}
\centering
 \includegraphics[width=\linewidth]{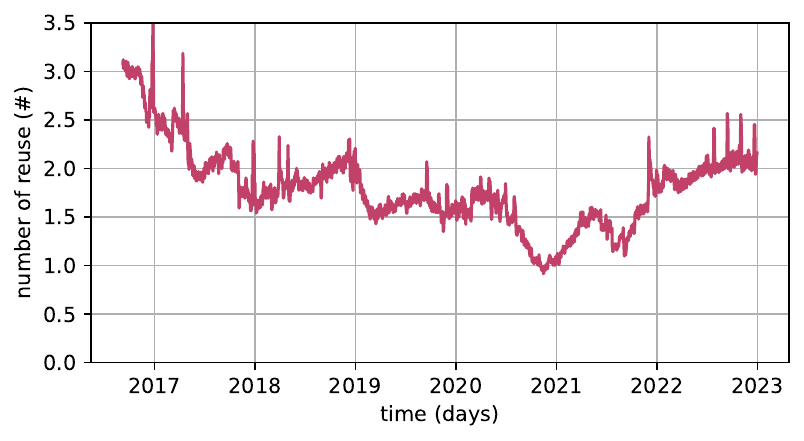}
\caption*{\footnotesize{Aggregate daily number of collateral re-use for reporting banks in MMSR, measuring the length 
of collateral chains. Only transactions between deposit-taking corporations are considered, all types of collateral are included. Source: Money Market Statistical Reporting database (MMSR).}}
 \caption{Length of the chain of collateral among the 47 largest banks of the eurozone} 
\label{fig:results/secured/transaction_view/collateral_re-use/collateral_re-use}
\end{figure}

As reported by \citet{LeCozEtAl-2024c}, various levels of collateral re-use, ranging from $0.1$ to $3$ have been measured across time and regions: notably a re-use rate around $1$ was observed in European money markets \citep{KellerEtAl-2014, EuropeanSystemicRiskBoard.-2017}, $0.6$ in Australia \cite{CheungEtAl-2014}, $0.1$ in Switzerland \citep{FuhrerEtAl-2016}, and $3$ in the US \citep{SCAGGS-2018}. We confirm a re-use rate around $1$ for the eurozone in Fig.~\ref{fig:results/secured/transaction_view/collateral_re-use/collateral_re-use} by measuring the weighted number of times the ISIN code of a given collateral appears in the banking system on a given day.

\section{The interbank network topology}

\paragraph{Sparse core periphery structure?} 
We define a link in the interbank network as the presence of at least one repo exposure between two banks over a specified aggregation period, which can range from one day to one year. Historically, interbank market networks exhibited low density and a core-periphery structure \citep{BechMonnet-2016,BlasquesEtAl-2018,Vari-2020, BossEtAl-2004}. In this configuration, a central 'core' of highly interconnected nodes is surrounded by a 'periphery' of less connected nodes that primarily link to the core rather than to each other.

\begin{figure}
\centering
 \includegraphics[width=\linewidth]{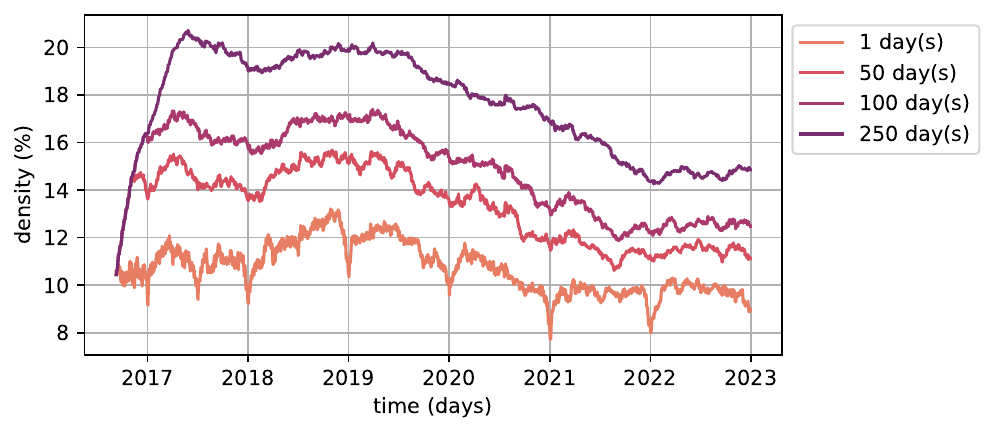}
 \caption*{\footnotesize{Density of the secured segment of the interbank money markets in the euro zone. A link between two reporting banks is defined as the existence of at least one repo transaction over different aggregation periods, each corresponding to a different color: (1) over 1 day, (2) over 50 days, (3) over 100 days and (4) over 250 days. Only transactions between deposit-taking corporations are considered, all types of collateral are included. Source: Money Market Statistical Reporting database (MMSR).}}
 \caption{Density of the repo interbank markets among the 47 largest banks of the eurozone}
\label{fig:results/secured/exposure_view/network_density}
\end{figure}

The switch of these markets towards secured transactions led, according to the MMSR data, to an increased network density. Indeed, Fig.~\ref{fig:results/secured/exposure_view/network_density} shows a network density ranging from $10\%$ to $20\%$ deepening on the link definition. We assume this higher density is due to the longer transaction maturity. The limited number of banks in our sample (47) prevented us from studying the core-periphery structure of secured markets.

\begin{figure}
\centering
 \includegraphics[width=\linewidth]{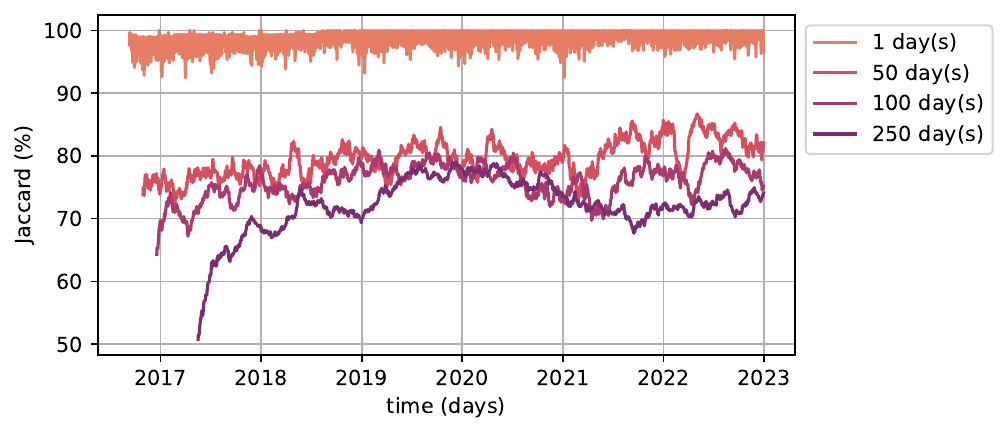}
\caption*{\footnotesize{Jaccard similarity coefficient of the secured segment of the interbank money markets in the euro zone. A link between two reporting banks is defined as the existence of at least one repo transaction over different aggregation periods, each corresponding to a different color: (1) over 1 day, (2) over 50 days, (3) over 100 days and (4) over 250 days. Only transactions between deposit-taking corporations are considered, all types of collateral are included. Source: Money Market Statistical Reporting database (MMSR).}}
 \caption{Stability of the repo interbank markets among the 47 largest banks of the eurozone}
\label{fig:results/secured/exposure_view/jaccard_index}
\end{figure}

\paragraph{Stable bilateral relationships}
The existence of stable interbank relationship lending has been documented by, among others, \citet{Furfine-1999,AfonsoEtAl-2013,BlasquesEtAl-2018}. We confirm this result in the case of secured markets by measuring the share of stable links from one period to another, namely the Jaccard network similarity index \citep{VermaAggarwal-2020}. Figure\ref{fig:results/secured/exposure_view/jaccard_index} shows a Jaccard network similarity index ranging from $80$ to $100\%$ depending on the aggregation period.

\paragraph{Asymmetric in and out degrees}
Several authors reported an asymmetry between in and out degree within unsecured interbank lending networks \citep{CraigvonPeter-2014,AnandEtAl-2015,Lux-2015}. Notably, \citet{CraigvonPeter-2014,AnandEtAl-2015} observe that banks in Germany have in general fewer lenders than borrowers.

\begin{figure}
\centering
 \includegraphics[width=0.7\linewidth]{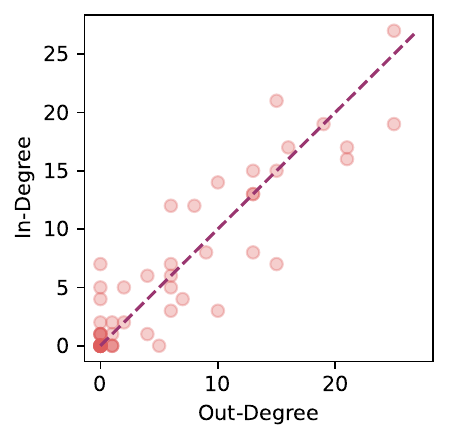}
 \caption*{\footnotesize{This figure presents the relationship between in and out-degree within the secured interbank money market segment in the eurozone, as of July 3rd 2022. Links are defined through the aggregation of transactions that occurred within the last $50$ days. Only transactions between deposit-taking corporations are considered, all types of collateral are included. Source: Money Market Statistical Reporting database (MMSR).}}
 \caption{In--degree as a function of out--degree on 31st December 2022}
\label{fig:results/in_out_degree_scatter_empirical}
\end{figure}

We observe a more symmetrical pattern in the case of the repo exposures among the 47 largest banks of the eurozone. Figure~\ref{fig:results/in_out_degree_scatter_empirical} shows that in--degrees expressed as a function of out-degree are almost symmetrical.

\section{Conclusion}
In this study, we analyzed secured transactions documented in the Money Markets Statistical Reporting database of the European Central Bank to investigate various characteristics of the interbank market among the 47 largest banks in the eurozone. Our findings reveal a significant increase in the volume of evergreen repurchase agreements coinciding with the implementation of the LCR regulation. Additionally, our measurement of collateral re-use rates aligns with existing literature on the subject. When examining the structure of the interbank network, our results confirm its high level of stability. However, we also identified a higher network density and more pronounced symmetry in in-degree and out-degree connections compared to unsecured markets. These insights contribute to a deeper understanding of the current dynamics and structural properties of the eurozone's interbank market.

\section{Acknowledgments}
We are grateful to the Financial Research Division of the Directorate General Research of the European Central for allowing us to access the MMSR database. We are indebted to Stefano Corradin, who contributed to our research through fruitful discussions. Finally, we thank Bertrand Hassani and the ANRT (CIFRE number 2021/0902) for providing us with the opportunity to conduct this research at Quant AI Lab. 

This research was conducted within the Econophysics \& Complex Systems Research Chair, under the aegis of the Fondation du Risque, the Fondation de l’École Polytechnique, the École Polytechnique and Capital Fund Management.

\FloatBarrier

\bibliographystyle{apsrev4-2} 
\bibliography{zotero}

\newpage

\appendix
\section{Data sources} \label{Data sources}
Our empirical analyses rely on the Money Market Statistical Reporting (MMSR) database which contains the transaction statistics of both the secured and unsecured legs of the euro area money markets. From July 2016 to December 2022 the database contains transaction-by-transaction level data on the euro money markets reported by a sample of 47 euro area banks\footnote{The full list of reporting agents is available on the \href{https://www.ecb.europa.eu/stats/financial_markets_and_interest_rates/money_market/html/index.en.html}{Money market statistical reporting webpage}}. They are required to report all their transactions of maturity inferior to one year with other financial institution\footnote{Formally it is legal entities of 10 sectors of the economy: general government and other government institutions, non-financial corporations, central banks, deposit taking corporations, money market funds, investment funds, insurance corporations, pension funds and other financial intermediaries - this includes financial auxiliaries, captive financial institutions and money lenders, as well as another bundled category called "other financial intermediaries, except insurance corporations and pension funds"}. On both segments, we have access to information on the type of money market instrument, the dates of trade, start and termination of the contract, the interest rate of the transaction, the volume, the sector of the counterparty, the volume of the transaction, the maturity of the transaction and the direction of the trading (lending vs borrowing). We also have the identifiers of the reporting banks and identifiers of the counterparties. For secured transactions, we also can identify the ISIN of the collateral. Banks in MMSR have the obligation to report all transactions until they mature, so that any deposit from another entity that is reported in the balance sheet of a given bank is reported everyday until the entity withdraws its deposits either partly or entirely.  
\paragraph{General data retreatment.} We keep any interbank transactions between MMSR reporting agents for which we have non missing reported information on the transaction volume, the LEI of the reporting agent, the LEI of the counterparty agent, the ISIN code of the collateral and the direction of the transaction (borrowing vs lending). We remove all canceled transactions from the database - i.e. a canceled transaction corresponds to any transaction which was initially reported and was later canceled, but is not equivalent to a transaction which was approved by the bank and later matured (these are the transactions we keep for our analysis). Additionally, we make sure that the dates in our sample only correspond to official euro area trading calendar dates\footnote{We list all calendar dates reported on the official statistical data warehouse of the ECB from the reported EONIA/ESTR rates time series.}.
\paragraph{Identification of evergreen repos.} Our main data transformation consists in extracting information on the evergreen repos from the MMSR database. Evergreen repos are repos with an infinite maturity but are not flagged as such in the MMSR database. Evergreens have a notice period that usually vary between 1 to 100 days and are reported as a repo transaction with the same maturity band everyday until the day when one or both trading counterparties decide to stop the transaction. That day, the two counterparties have to agree on a date of final maturity, which will then be reported in the database. We thus identify evergreens as any repeated transactions between the two same counterparties for at least 1 day. Specifically, we identify a unique repo as the combination of the two identifiers of the transacting counterparties (the lender and the borrower), the nominal amount and the maturity and the ISIN code of the transaction. Any repeated combination of these unique repos for more than a day are considered evergreens. 
\end{document}